\documentclass[aps,prd,preprint,showpacs,floats,nofootinbib]{revtex4-1}

\usepackage[dvips]{graphicx}
\usepackage{subfigure}
\usepackage{amssymb}
\usepackage{amsmath}
\usepackage{bm}
\usepackage{times}
\usepackage{epsfig}
\usepackage{color}

\def\beq{\begin{equation}}
\def\eeq{\end{equation}}
\def\be{\begin{equation}}
\def\ee{\end{equation}}
\def\bea{\begin{eqnarray}}
\def\eea{\end{eqnarray}}
\def\to{\rightarrow}

\begin{document}
\title{%%Impact of annihilation amplitudes in charmless B decays on new physics searches
        Flavor dependence of annihilation parameters in QCD factorization}
\author{Kai Wang}
\email{wangkai1@zju.edu.cn}
\author{Guohuai Zhu}
\email{zhugh@zju.edu.cn}
\affiliation{Zhejiang Institute of Modern Physics and Department of Physics, Zhejiang University, Hangzhou, Zhejiang 310027, CHINA}

\begin{abstract}
For $B_{d,s} \to \pi^\mp K^\pm$ and $K^{(\ast)} K^{(\ast)}$ decays, the flavor symmetry breaking effects may be particularly small since the final state interactions should be the same
between the corresponding $B_d$ and $B_s$ decays due to the charge conjugation symmetry of the final states. This is consistent with the newly measured direct CP asymmetry of $B_s \to \pi^+ K^-$. These decays are thus supposed to be important in testing the Standard Model and in probing new physics effects. However, the observation of pure annihilation decay $B_s \to \pi^+ \pi^-$ appears to imply a large annihilation scenario with $\rho_A \sim 3$, in contrast to the case of $\rho_A \sim 1$ in $B_{u,d}$ decays in the framework of QCD factorization. This seems to indicate unexpectedly large flavor symmetry breaking effects between the annihilation amplitudes of $B_s$ and $B_{u,d}$ decays. This apparent contradiction could be resolved by noticing that there is a priori no reason to justify the common practice of assuming the universality of annihilation parameters for different Dirac structures of effective operators. We then argue that, for $B_{d,s} \to \pi^\mp K^\pm$ decays, the flavor symmetry breaking effects of annihilation amplitudes have all been included in the initial state decay constants and are thus small. But the flavor symmetry breaking effects in $B_{d,s} \to K^{(\ast)} K^{(\ast)}$ decays are likely to be much larger, as part of the annihilation topologies of $B_s \to K K$ decay could be related to $B_s \to \pi^+ \pi^-$ decay. Therefore when new physics effects are searched for in these decay channels, care must be taken to consider the potentially large flavor symmetry breaking effects in more details.
\end{abstract}

\maketitle

\section{Introduction}
Charmless hadronic B decays, and in particular their CP asymmetries, are very sensitive to new physics
since the decay amplitudes are either highly Cabibbo-suppressed or loop suppressed in the Standard Model (SM).
However, it is notoriously difficult to calculate the amplitudes of hadronic B decays reliably, due to non-perturbative QCD interactions. These amplitudes are usually evaluated using factorization methods, which however are only valid to the leading order of power expansion in $1/m_b$.
To go beyond the leading power, model dependence may enter. Therefore, in many cases, it is hard to distinguish new physics signal from the SM backgrounds. For example, the difference of direct CP asymmetries for $B^0 \to \pi^- K^+$ and $B^+ \to \pi^0 K^+$ is observed to be $-0.126 \pm 0.022$ \cite{Amhis:2012bh}, which is unexpectedly large since it would vanish in the limit of isospin symmetry. This so-called $B \to K\pi$ CP puzzle, as first discovered by the Belle collaboration \cite{Lin:2008zzaa}, might imply new physics in the electroweak penguin sector which violates isospin symmetry. However, a mundane explanation of large color-suppressed tree amplitude due to non-perturbative QCD is at least equally possible (see, for example, \cite{Cheng:2009cn} and references therein).

Flavor symmetry is a powerful tool in heavy flavor physics. It has been implemented over the last two decades to study the CP violating relations and annihilation contributions in charmless B decays (see for example \cite{Deshpande:1994ii,He:1998rq}). Generally, SU(3) flavor symmetry may receive large corrections at about $20\%$ level, except isospin which is a good symmetry at a few percent level. But the flavor symmetry breaking effects could be much smaller in some cases. For example, Lipkin \cite{Lipkin:2005pb} noticed that, for $B_{d,s} \to \pi^\mp K^\pm$ decays, the U-spin ($d\leftrightarrow s$) symmetry breaking effects should be unusually small since the strong phases from final state interactions are exactly the same due to the charge conjugation symmetry of the final states. Therefore it could be a robust test of the SM vs New Physics to check the relation between the
direct CP asymmetries of these two decay channels. Interestingly, direct CP asymmetry of $B_{s} \to \pi^+ K^-$ has been measured very recently by the LHCb collaboration to be $0.27 \pm 0.04 \pm 0.01$ \cite{Aaij:2013iua}, which is the first observation of CP violation in $B_s$ decays. This measurement is consistent with the SM relation between direct CP asymmetries of $B_{d,s} \to \pi^\mp K^\pm$ decays. It has also been shown that, by a combined use of flavor symmetries and factorization method,  $B_{d,s} \to K^{(\ast)} K^{(\ast)}$ decays \cite{Descotes-Genon:2006wc,Datta:2006af,DescotesGenon:2007qd,Ciuchini:2007hx,DescotesGenon:2011pb,Bhattacharya:2012hh,Ciuchini:2012gd} may play an important role in testing the SM and in probing new physics effects.

However, as we will see in the following, recent measurements on pure annihilation decays $B_s \to \pi^+ \pi^-$ and $B_d \to K^+ K^-$ may indicate
significant violation of flavor symmetry in $B_{d,s}$ decays, at least for the annihilation amplitudes.
The first evidence of $B_s \to \pi^+ \pi^-$ decay was reported by the CDF collaboration \cite{Aaltonen:2011jv} to be
\begin{align}
{\cal B}(B_s \to \pi^+ \pi^-)&=(0.57 \pm 0.15 \pm 0.10) \times 10^{-6}~,
\end{align}
where the first errors are statistical and the second systematic. It was soon confirmed by the LHCb collaboration with $0.37~\mbox{fb}^{-1}$ data \cite{Aaij:2012as} as
\begin{align}
{\cal B}(B_s \to \pi^+ \pi^-)&=(0.95^{+0.21}_{-0.17} \pm 0.13) \times 10^{-6}~.
\end{align}
The average of the above measurements gives $(0.73 \pm 0.14) \times 10^{-6}$ \cite{Amhis:2012bh}. One expects the branching ratio of $B_d \to K^+ K^-$ should not be very different from that of $B_s \to \pi^+ \pi^-$ as they can be related to each other by U-spin symmetry. But the experimental efforts of the CDF and LHCb collaborations reveals a surprisingly small result \cite{Amhis:2012bh}
\begin{align}
{\cal B}(B_d \to K^+ K^-)&=(0.12 \pm 0.06) \times 10^{-6}~,
\end{align}
which is several times smaller than the branching ratio of $B_s \to \pi^+ \pi^-$ and may imply unexpectedly large flavor symmetry breaking effects.
As better understanding on the flavor symmetry breaking is crucial to separate new physics signal from the SM contributions, we will reinvestigate the flavor symmetry breaking effects in annihilation amplitudes, which is important for charmless hadronic B decays.

The potential importance of weak annihilation amplitudes was noticed first in \cite{Keum:2000ph,Keum:2000wi,Lu:2000em} for charmless B decays and was predicted in perturbative QCD method in \cite{Ali:2007ff,Li:2004ep,Li:2005vu}. Although being formally power suppressed in $\Lambda_{QCD}/m_b$ in QCD factorization method (QCDF) \cite{Beneke:1999br,Beneke:2000ry,Beneke:2001ev,Beneke:2003zv}, weak annihilation contributions are supposed to be important, together with the chirally-enhanced power corrections, to account for the large branching ratios and CP asymmetries of penguin-dominated B decays. In soft collinear effective theory \cite{Bauer:2000yr,Bauer:2001cu,Bauer:2004tj}, it was argued in \cite{Jain:2007dy,Arnesen:2006vb} that annihilation contributions are factorizable and may not be significant numerically in charmless B decays. Inspired by the experimental progress, there are some theoretical interest \cite{Zhu:2011mm,Xiao:2011tx,Chang:2012xv,Gronau:2012gs} recently in these pure annihilation decays. The authors of \cite{Xiao:2011tx} calculated $B_s \to \pi^+ \pi^-$ and $B_d \to K^+ K^-$ decays in perturbative QCD method with the results to be in agreement with the experimental data. Refs. \cite{Zhu:2011mm,Chang:2012xv} investigated these channels, together with other charmless hadronic B decays, in QCDF and found that SU(3) breaking effects should be taken into account for annihilation parameters. In addition, a large annihilation scenario seems to be favored in $B_s$ decays. While the authors of \cite{Gronau:2012gs} discussed the possibility that these decays can also be generated by rescattering from processes such as color-favored tree amplitudes.

This paper is organized as follows. In the next section, we shall discuss in detail the possible flavor dependence of annihilation parameters for charmless B decays to two light pseudoscalar mesons in QCDF method. We then conclude with a summary in section III.

\section{Annihilation amplitudes in QCD factorization}

\begin{figure}
\includegraphics[width=11cm]{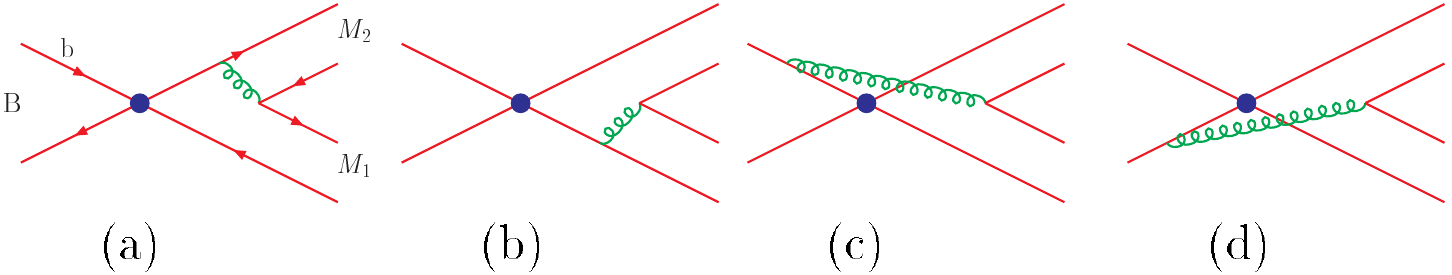}\caption{\label{fig:anni1} Annihilation contributions at leading order of $\alpha_s$. }
\end{figure}
We first briefly review the annihilation amplitudes of $B \to PP$ decays in QCDF method, one may refer to \cite{Beneke:2001ev,Beneke:2003zv} for the details. The effective Hamiltonian of $\Delta B=1$ can be expressed as
\begin{align}
{\cal H}_{eff}=\frac{G_F}{\sqrt{2}}\sum_{i=1}^{10} \sum_{p=u,c} \lambda_{pq} C_i(\mu) Q_i^q (\mu)~,
\end{align}
where $\lambda_{pq}=V_{pb}^\ast V_{pq}$ ($q=d$ or $s$) is a CKM factor, $C_i (\mu)$ is the Wilson coefficient
which is perturbatively calculable from first principles. The four-quark effective
operators $Q^q_{1,2}$, $Q^q_{3,...,6}$ and $Q^q_{7,...,10}$ are tree level, QCD penguin and electroweak penguin
operators, respectively. These effective operators can contribute to the annihilation amplitudes, as shown in Fig. \ref{fig:anni1}. The basic build blocks for pseudoscalar final states may be simplified by taking asymptotic light-cone distribution amplitudes and the approximation $r_\chi^\pi \simeq r_\chi^K \equiv r_\chi$
\begin{align}
&A_1^i \simeq A_2^i \simeq 2\pi \alpha_s \left ( 9(X_A-4+\pi^2/3)+r_\chi^2 X_A^2\right )~, \hspace*{0.3cm} A_3^i\simeq 0~, \nonumber \\
&A_1^f=A_2^f=0~, \hspace*{0.3cm} A_3^f \simeq 12\pi \alpha_s r_\chi (2X_A^2-X_A)~.
\end{align}
In the first (last) two diagrams of Fig. \ref{fig:anni1}, the gluon is emitted from the final (initial) quarks. Correspondingly, their contributions to the basic building blocks are labeled by the superscript `f' (`i'). The subscripts 1,2,3 of $A_k^{i,f}$ denote the different Dirac structure of the four-quark operators as $(V-A)(V-A)$, $(V-A)(V+A)$ and $(S-P)(S+P)$, respectively.
The ratio $r_\chi$ is defined by $r_\chi=2m_K^2/(m_b(m_q+m_s))$ with $m_q$ the average of the up and down quark masses.
$X_A$ parameterizes the endpoint singularity as
\begin{align}
X_A=\ln \frac{m_B}{0.5 \mbox{GeV}}\Big(1+\rho_A e^{i\phi_A} \Big)~.
\end{align}
Notice that $\phi_A$ is an arbitrary strong phase and normally $\rho_A \sim 1$ is assumed, which reflects our ignorance on the
annihilation amplitudes dominated by the soft gluon interaction. In principle, $X_A$ may vary not only for different initial and final states but also for different Dirac structure of the effective operators. Since different initial and final states can be related to each other by flavor symmetry, the annihilation parameters $\rho_A$ and $\phi_A$ should only vary mildly for different decay channels. However, there is no a priori reason for the annihilation parameters to be the same for $A_k^{i,f}$ with different subscript, though in practice they were taken to be universal for simplicity. As we shall see in the following, current experimental data may indicate that a universal set of annihilation parameters for $A_1^{i}$, $A_2^i$ and $A_3^f$ are disfavored.

It is convenient to define further the annihilation coefficients b's as
\begin{align} \label{eq:bi's}
b_1&=\frac{C_F}{N_c^2}C_1 A_1^i~, \hspace*{1cm}  b_3=\frac{C_F}{N_c^2} \Big [ C_3 A_1^i + (C_5+N_c C_6) A_3^f \Big ]~, \nonumber \\
b_2&=\frac{C_F}{N_c^2}C_2 A_1^i~, \hspace*{1cm}  b_4=\frac{C_F}{N_c^2} \Big [ C_4 A_1^i + C_6 A_2^i \Big ]~, \nonumber \\
b_{3,EW}&=\frac{C_F}{N_c^2} \Big [ C_9 A_1^i + (C_7+N_c C_8) A_3^f \Big ]~,\hspace*{1cm} b_{4,EW}=\frac{C_F}{N_c^2} \Big [ C_{10} A_1^i + C_8 A_2^i \Big ]~.
\end{align}
Numerically, $b_{3,EW}$ and $b_{4,EW}$ are negligible due to small Wilson coefficients. For the rest of b's, only $b_3$ contains $A_3^f$ term, which is actually a dominant one if the magnitude of $A_1^i$ are not much larger than that of $A_3^f$. This is a key observation for our later analysis.

\begin{figure}
\includegraphics[scale=0.8]{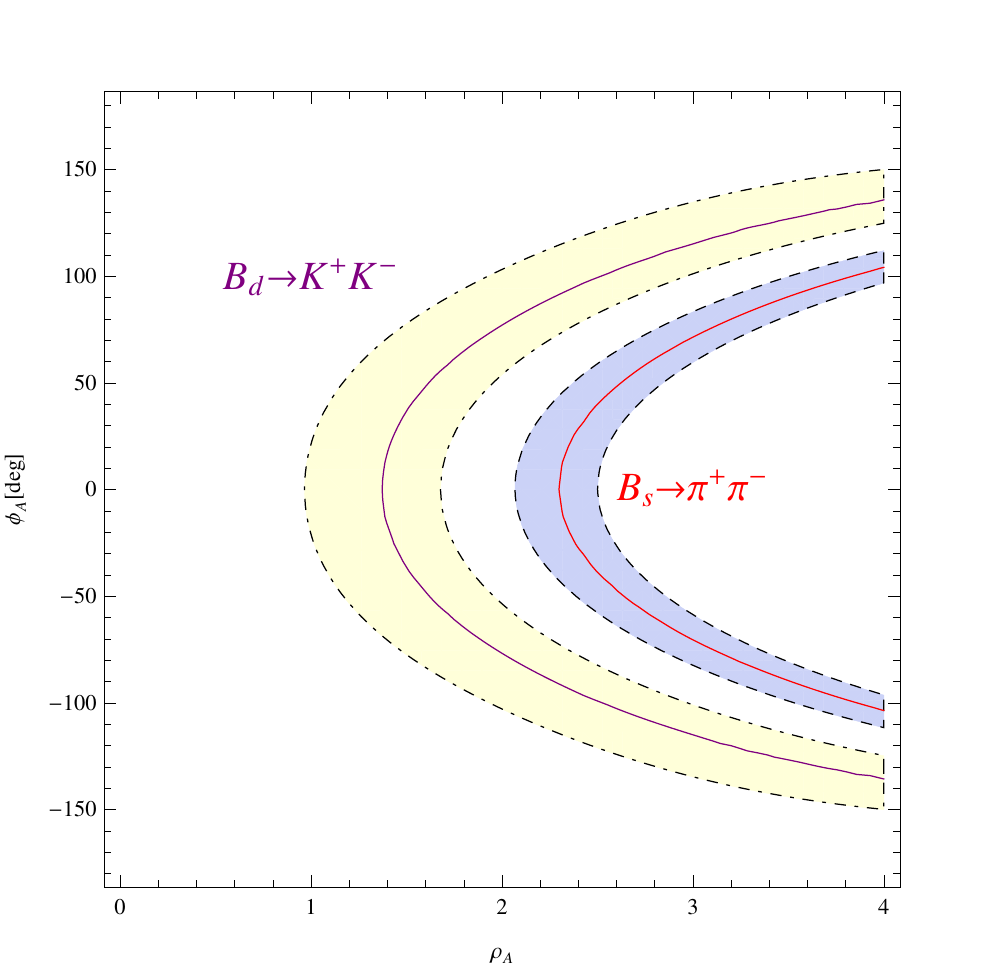} \caption{\label{fig:pipiKK} Contour plot of the branching ratios of $B_s \to \pi^+ \pi^-$ and $B_d \to K^+ K^-$ decays as functions of  the annihilation parameters $\rho_A$ and $\phi_A$. The solid lines represent the experimental central values and the grey regions correspond to one sigma contour. }
\end{figure}

As discussed in \cite{Zhu:2011mm,Chang:2012xv}, the recent experimental measurements have revealed that the scenario of universal annihilation parameters for all $B \to PP$ decays is in somewhat disagreement with pure annihilation decays $B_s \to \pi^+ \pi^-$ and $B_d \to K^+ K^-$. This point can be seen clearly in Fig. \ref{fig:pipiKK}, where there is no overlap between the regions of annihilation parameters favored by these two decays.

To draw Fig. \ref{fig:pipiKK}, we have taken the following input parameters \cite{Beringer:1900zz}
\begin{align}\label{eq:input1}
f_{B_s}=230~\mbox{MeV}~, \hspace*{0.5cm} f_{B}=190~\mbox{MeV}~, \hspace*{0.5cm} m_b(m_b)=4.2~\mbox{GeV}~, \hspace*{0.5cm} m_s(2~\mbox{GeV})=95~\mbox{MeV}~,
\end{align}
and the Wolfenstein parameters \cite{Charles:2004jd}
\begin{align}\label{eq:CKM}
A=0.812~, \hspace*{1cm} \lambda=0.2254~, \hspace*{1cm} \bar{\rho}=0.144~, \hspace*{1cm} \bar{\eta}=0.342~.
\end{align}
It is then straightforward to evaluate the decay amplitudes of $B_s \to \pi^+ \pi^-$ and $B_d \to K^+ K^-$, which can be expressed in QCDF as \cite{Beneke:2003zv}
\footnote{It has been pointed out in \cite{DeBruyn:2012wj,DeBruyn:2012wk} that a correction factor due to sizable decay width difference in $B_s$ system has to be included when
the experimentally measured branching ratios are compared to the theoretical branching ratios. For flavor-specific decay such as $B_s \to \pi^+ K^-$, the correction factor is about $1\%$ and can be neglected. But for the cases of $B_s \to \pi^+ \pi^-$, $K^+ K^-$, the correction factor could be as large as up to $10 \%$. However more experimental information such as time-dependent analysis is required to determine the correction factor, so we shall not consider this effect in the following.}
\begin{align}\label{eq:amp-pipiKK}
A(B_s \to \pi^+ \pi^-)&=B^s_{\pi\pi}\left (V_{ub}^\ast V_{us}\Big [ b_1+2b_4+\frac{1}{2}b_{4,EW}\Big ]+ V_{cb}^\ast V_{cs}\Big[ 2b_4+\frac{1}{2}b_{4,EW}\Big ] \right ) \nonumber \\
A(B_d \to K^+ K^-)&=B^d_{KK}\left (V_{ub}^\ast V_{ud}\Big [ b_1+2b_4+\frac{1}{2}b_{4,EW}\Big ]+ V_{cb}^\ast V_{cd}\Big[ 2b_4+\frac{1}{2}b_{4,EW}\Big ] \right )
\end{align}
with
\begin{align}
B^s_{\pi\pi}=i \frac{G_F}{\sqrt{2}} f_{B_s} f_\pi f_\pi~,\hspace*{1cm} B^d_{KK}=i \frac{G_F}{\sqrt{2}} f_{B} f_K f_K~.
\end{align}

As only experimental uncertainties are included in Fig. \ref{fig:pipiKK}, one may wonder whether the situation may change when theoretical
uncertainties are considered. From Eq. (\ref{eq:amp-pipiKK}), it is clear that the theoretical uncertainties of the ratio ${\cal B}(B_s \to \pi^+ \pi^-)/{\cal B}(B_d \to K^+ K^-)$ involves only $f_{B_s}/f_B$ and the annihilation parameters, while the CKM dependence is almost canceled. As the lattice QCD calculations have obtained impressive results of $f_{B_s}/f_B=1.201\pm 0.017$ \cite{Laiho:2009eu} as an average with small errors, the inclusion of theoretical uncertainties would not change our conclusion about the failure of the universal annihilation parameters.

\begin{figure}
\includegraphics[scale=0.8]{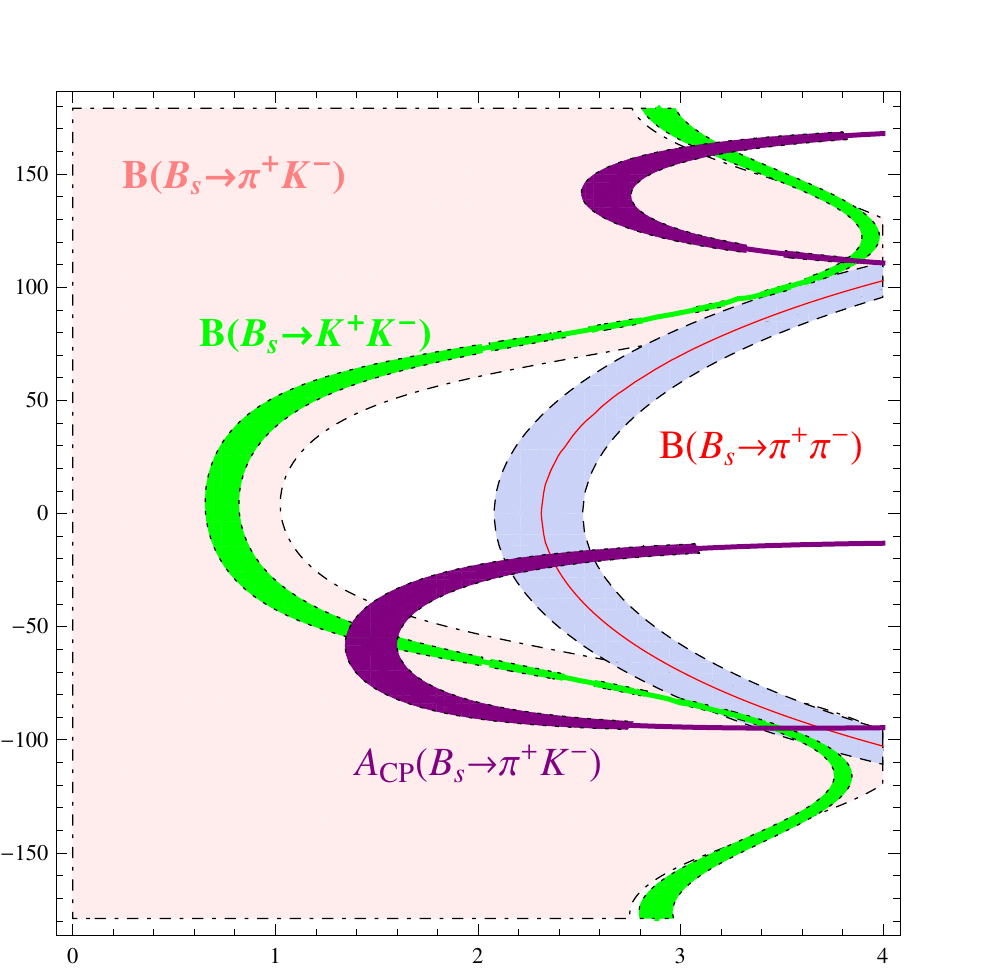} \caption{\label{fig:Bs-all} Contour plot of the branching ratios of $B_s \to \pi^+ \pi^-$, $\pi^+ K^-$, $K^+ K^-$ decays and direct CP asymmetry of $B_s \to \pi^+ K^-$ as functions of the annihilation parameters $\rho_A$ and $\phi_A$. The grey (colored) regions correspond to one sigma contour. }
\end{figure}

Considering the SU(3) flavor symmetry breaking, it is not a surprise at all that the scenario of universal annihilation parameters does not work.
Actually, it has long been assumed (see, for example, Refs \cite{Beneke:2003zv,Cheng:2009cn,Cheng:2009mu}) that the annihilation parameters are slightly different between $B_{u,d}$ and $B_s$ decays. Following this assumption, the annihilation parameters have been carefully studied in \cite{Chang:2012xv}, which implied large annihilation corrections of $\rho_A \sim 3$ for $B_s$ decays, in contrast to the case of $\rho_A \simeq 1$ widely used before. Very recently, first observation of direct CP violation in charmless Bs decays, $A_{CP}(B_s \to \pi^+ K^-)$, has been reported by LHCb \cite{Aaij:2013iua} to be $0.27 \pm 0.04(stat) \pm 0.01(syst)$. The CDF collaboration also reported an evidence of the CP violation to be $0.22 \pm 0.07(stat) \pm 0.02(syst)$ \cite{CDFnote}. A naive average of the latest results yields $0.26\pm0.04$. Including this significantly improved data, we confirm that large annihilation scenario for $B_s \to PP$ decays is still consistent with all the experimental data, as shown in Fig. \ref{fig:Bs-all}. The overlap regions in the figure represent that there exist two solutions satisfying all the constraints: one with $\rho_A \simeq 3.5$, $\phi_A \simeq -100^\circ$ and the other with $\rho_A \simeq 3.8$, $\phi_A \simeq 110^\circ$.

To draw Fig. \ref{fig:Bs-all}, we have adopted the form factor $F^{B_sK}=0.24$ as suggested by \cite{Cheng:2009mu}. One can then easily fit the experimental data of $10^6 {\cal B}(B_s \to \pi^+ K^-)=5.4\pm 0.6$ \footnote{Unless stated otherwise, we shall always cite Heavy Flavor Averaging Group \cite{Amhis:2012bh} for the experimental data.}, though it is nontrivial to fit the data of $10^6 {\cal B}(B_s \to K^+ K^-)=24.5\pm 1.8$. Therefore ${\cal B}(B_s \to \pi^+ K^-)$ only gives a rather weak constraint on $(\rho_A,~\phi_A)$, which is plotted as the light pink region in the figure. Our results are not very sensitive to the parameters of wave functions, so we simply take \cite{Beneke:2003zv,Ball:2006wn}
\begin{align}\label{eq:other-input}
\lambda_B=200~\mbox{MeV}~,\hspace*{0.7cm} a_2^\pi=0.25~,\hspace*{0.7cm} a_1^K=0.06~,\hspace*{0.7cm} a_2^K=0.25~.
\end{align}

In a word, the large annihilation scenario for $B_s$ decays seems to be good, which is mainly required to fit the large ${\cal B}(B_s \to \pi^+ \pi^-)$. However, the relatively small ${\cal B}(B_d \to K^+ K^-)$ indicates that the case of $\rho_A \sim 1$ is good enough for $B_{u,d}$ decays.
This may suggest that the flavor symmetry breaking effects are unexpectedly large for annihilation amplitudes between $B_s$ and $B_{u,d}$ decays.

But as pointed out in \cite{Lipkin:2005pb}, the flavor symmetry breaking effects are probably exceptionally small in $B_{d,s} \to \pi^\mp K^\pm$ decays since the final state interactions should be exactly the same for these two decay channels. It is then straightforward to obtain a well-known relation between the direct CP violations of $B_{d,s} \to \pi K$ decays
\begin{align}\label{eq:piK-relation}
-\frac{A_{CP}(B_s \to \pi^+ K^-)}{A_{CP}(B^0 \to \pi^- K^+)} =
\frac{{\cal B}(B^0 \to \pi^- K^+)}{{\cal B}(B_s \to \pi^+ K^-)}\frac{\tau(B_s)((m_{B_s}^2-m_K^2)f_\pi F^{B_sK})^2}{\tau(B^0)((m_{B_d}^2-m_\pi^2)f_K F^{B\pi})^2}~,
\end{align}
which have included explicitly part of U-spin symmetry breaking effects in terms of decay constants and form factors. Experimentally, left hand side of Eq. (\ref{eq:piK-relation}) equals to $3.0 \pm 0.5$, while the right hand side equals to $2.2 \pm 0.6$ if we take the form factor $F^{B\pi}=0.26 \pm 0.03$ \cite{Ball:2004ye,Duplancic:2008ix} \footnote{A recent calculation \cite{Khodjamirian:2011ub} gives a slightly larger value of $0.28^{+0.02}_{-0.03}$. However since $B^0 \to \pi^+ \pi^-$ decay prefers smaller form factor (to be discussed later), we shall use the central value $0.26$ in our analysis.} estimated by light-cone sum rules. So Eq. (\ref{eq:piK-relation}) is consistent within roughly one sigma with the experimental measurements and there is no sign of large flavor symmetry breaking beyond decay constants and form factors in $B_{d,s} \to \pi^\mp K^\pm$ decays. This observation is in a sense disagree with the above-discussed scenario of large annihilation magnitude $\rho_A \ge 3$ in $B_s$ decays together with $\rho_A \sim 1$ in $B_{u,d}$ decays. This issue is also important in $B_{d,s} \to K^{0(\ast)} K^{0(\ast)}$ decays, which may play an important role in testing the SM and in probing new physics effects \cite{Descotes-Genon:2006wc,Datta:2006af,DescotesGenon:2007qd,Ciuchini:2007hx,DescotesGenon:2011pb,Bhattacharya:2012hh,Ciuchini:2012gd}.
It stimulates us to reinvestigate the annihilation amplitudes in QCDF and look for different possibility not discussed before.

Notice that, for annihilation part, only $b_3$ and $b_{3,EW}$ appear in $B_{d,s} \to \pi^\mp K^\pm$ decays. We have mentioned before that $b_{3,EW}$ is negligible due to small electroweak Wilson coefficients and $b_3$ is in general dominated by the $A_3^f$ term as can be checked from Eq. (\ref{eq:bi's}). Since $A_3^f$ represents the gluon emitted from the final quarks, i.e. the first two diagrams of Fig. \ref{fig:anni1}, the initial state dependence of the corresponding annihilation corrections must have all been included in decay constants. According to the definition, all the decay constants have been taken outside the building blocks $A_k^{i,f}$. Therefore, $A_3^f$ is independent of initial state and must be the same for $B_s$ and $B_d$ decays to the same final states. According to the above reasoning, the annihilation amplitudes of $B_{d,s} \to \pi^\mp K^\pm$ decays are roughly the same and we may take just one set of annihilation parameters $(\rho_A, \phi_A)$ for these two decays. This is also in agree with the observation that flavor symmetry breaking should be very small in $B_{d,s} \to \pi^\mp K^\pm$ decays, as noticed first in \cite{Lipkin:2005pb}. It is then straightforward to determine the annihilation parameters for $B_{d,s} \to \pi^\mp K^\pm$ decays.

\begin{figure}
\centering
\subfigure[]{
\label{fig:pikc}
\includegraphics[width=0.3\textwidth]{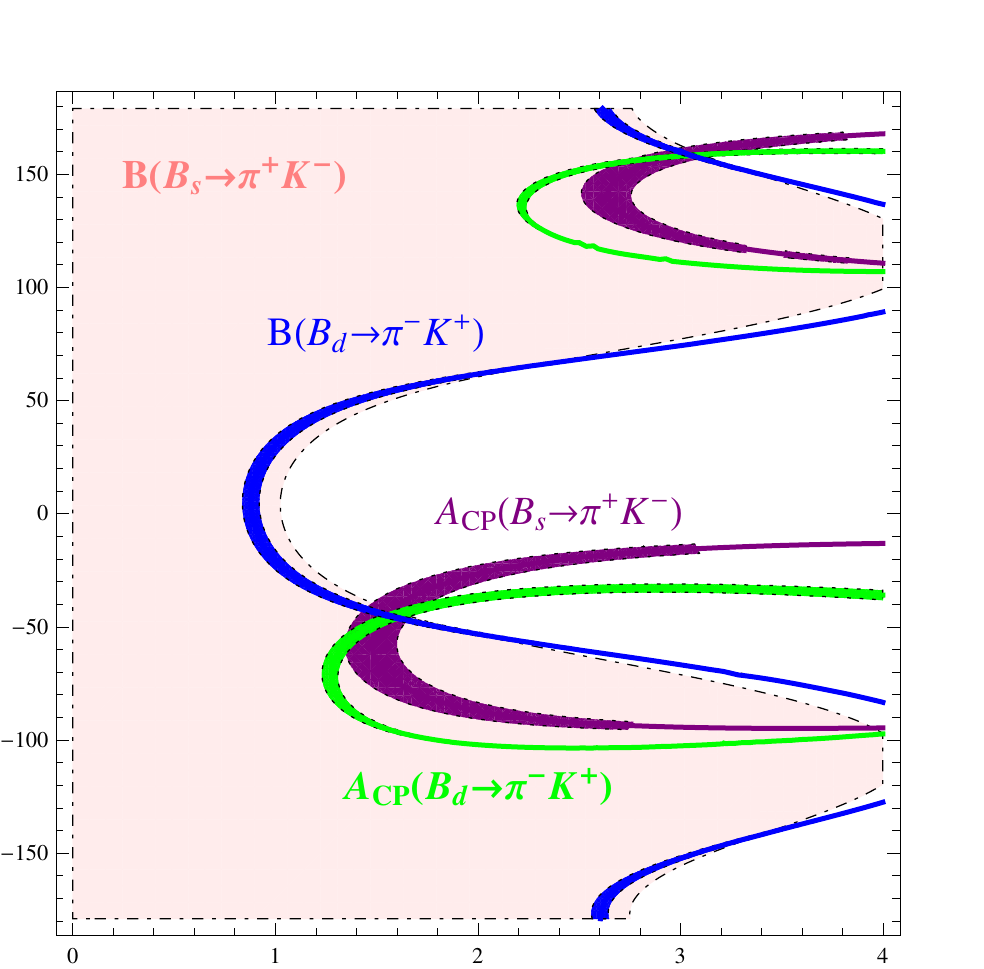}\hspace*{0.5cm}}
\subfigure[]{
\label{fig:pikb}
\includegraphics[width=0.3\textwidth]{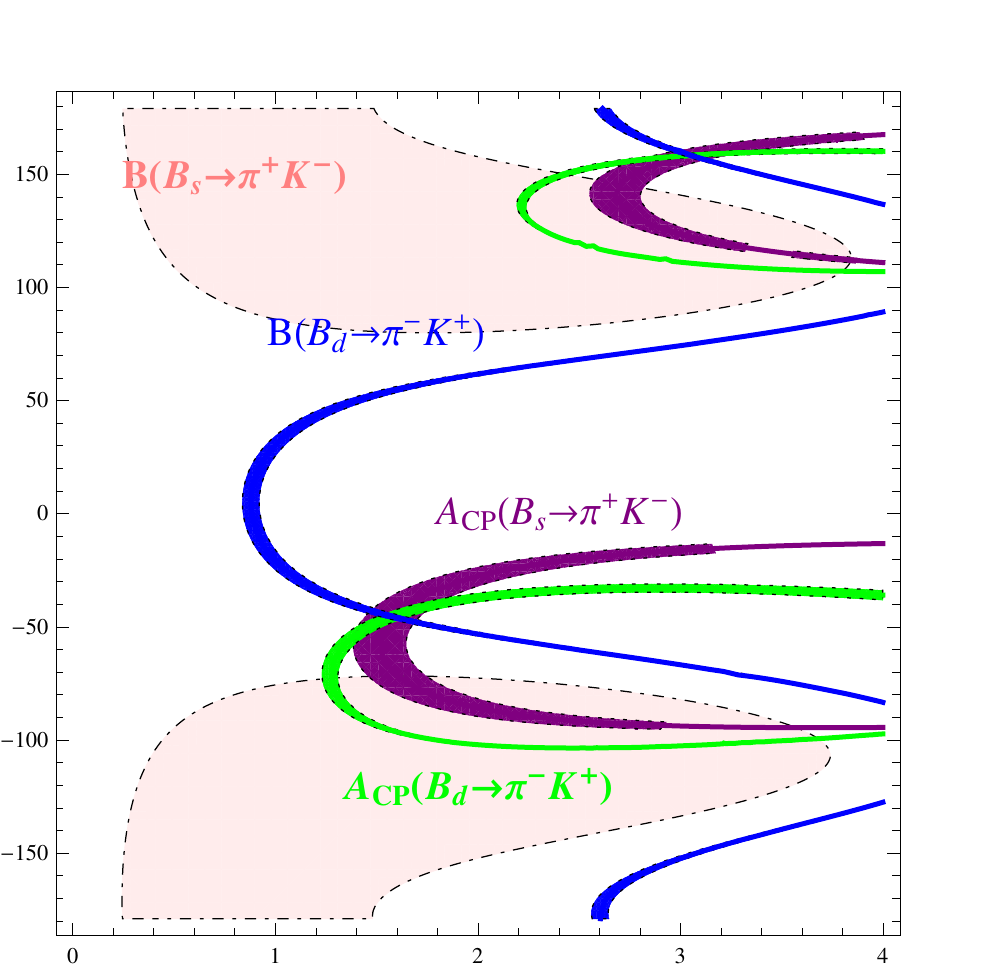}\hspace*{0.5cm}}
\subfigure[]{
\label{fig:piks}
\includegraphics[width=0.3\textwidth]{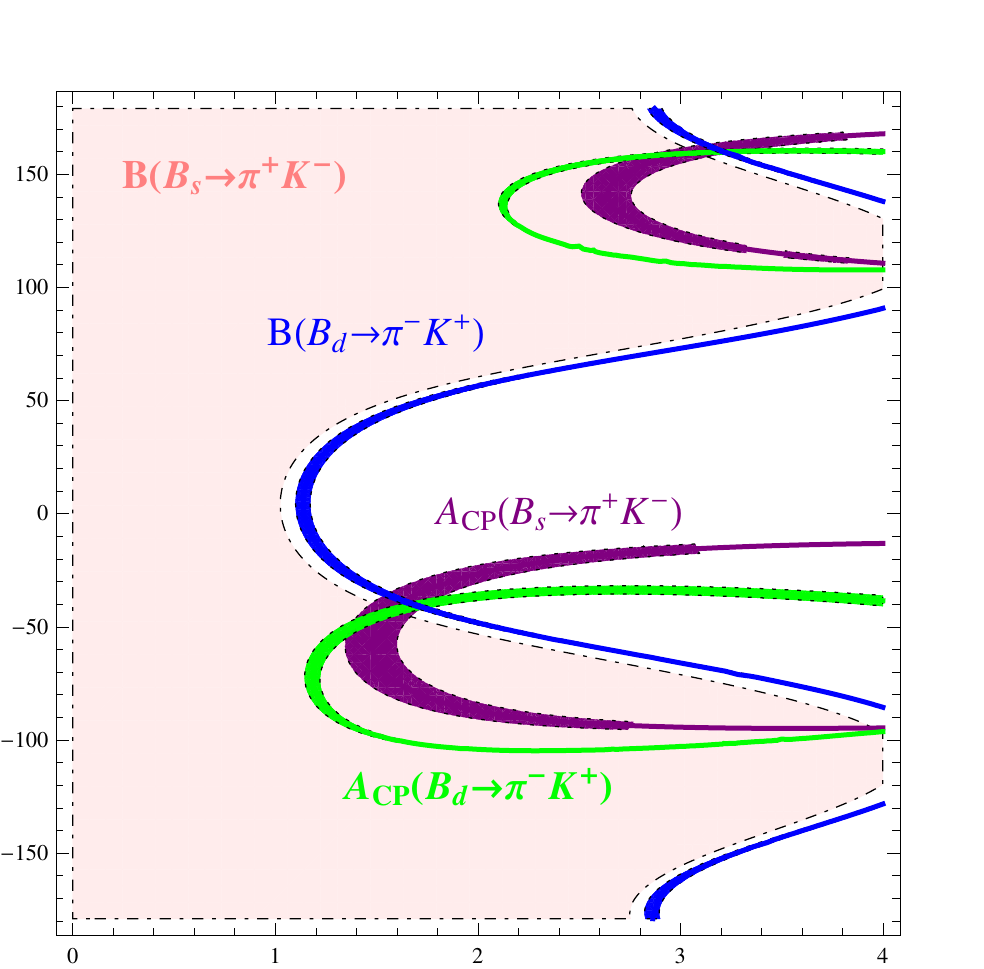}}
 \caption{\label{fig:pik} Contour plot of the branching ratios and direct CP asymmetries of $B_{d,s} \to \pi^\mp K^\pm$ decays as functions of the annihilation parameters $\rho_A$ and $\phi_A$. The form factors $F^{B_sK}=0.24$ and $F^{B\pi}=0.26$ are taken in the first figure. A slightly larger
 $F^{B_sK}=0.25$ is adopted in the second figure while $F^{B\pi}=0.23$ is used in the last figure. The grey (colored) regions correspond to one sigma contour with only experimental uncertainties included.}
\end{figure}

The small overlap regions in Fig. \ref{fig:pikc} shows clearly that one set of annihilation parameters are consistent with the experimental measurements of the branching ratios and direct CP asymmetries of $B_{d,s} \to \pi^\mp K^\pm$ decays. There appear two solutions in the plane of $(\rho_A, \phi_A)$, one around $(1.6,~-45^\circ)$ and the other around $(3.1,~158^\circ)$. But these two solutions actually correspond to the same value of annihilation coefficient $b_3$ in the decay amplitude. Therefore there is physically only one solution at the level of annihilation coefficients b's and one is free to choose either solution in the plane of $(\rho_A, \phi_A)$. For convenience, we shall take the solution of $(\rho_A, \phi_A)=(1.6,~-45^\circ)$ in the following.

As the theoretical inputs are fixed in Fig. \ref{fig:pikc}, it is interesting to discuss the numerical impacts on this solution by varying the inputs. For the form factor $F^{B_sK}$, QCD sum rules predicts $0.30^{+0.04}_{-0.03}$ \cite{Duplancic:2008tk}, but we choose $0.24$ in Fig. \ref{fig:pikc}. The reason is that a just slightly larger value of $F^{B_sK}=0.25$ would lead to no solution as shown in Fig. \ref{fig:pikb}, due to shrinkage of the pink region corresponding to the experimental constraint from ${\cal B}(B_s \to \pi^+ K^-)$. Actually the pink region would completely disappear for $F^{B_sK}>0.26$ with the CKM parameters fixed. This is because ${\cal B}(B_s \to \pi^+ K^-)$ is dominated by the tree-level amplitude which is basically determined by the factor $|V_{ub}|F^{B_sK}$. The CKM parameters in Eq. (\ref{eq:CKM}) determined by the CKMfitter Group corresponds to $|V_{ub}|=3.54 \times 10^{-3}$, which is consistent with the exclusive determination of $|V_{ub}|$, though a bit smaller than the inclusive determination of $|V_{ub}|$. But even with this relatively small $|V_{ub}|$, the experimental data on ${\cal B}(B_s \to \pi^+ K^-)$ still excludes $F^{B_sK}>0.25$, in contrast to the estimation of QCD sum rules. That is why we choose $F^{B_sK}=0.24$, following the choice of \cite{Cheng:2009mu}. A larger $|V_{ub}|$ would require an even smaller form factor of $F^{B_sK}$ to fit the experimental branching ratio of $B_s \to \pi^+ K^-$.

For $B_d \to \pi^- K^+$ decay, it depends on the form factor $F^{B\pi}$ which is estimated to be $0.26 \pm 0.03$ by light-cone sum rules \cite{Duplancic:2008ix,Khodjamirian:2011ub}. In Fig. \ref{fig:pik}, the cental value is used as default and one may vary $F^{B\pi}$ within the errors to check the impact on annihilation parameters. We have shown in Fig. \ref{fig:piks} with the case of $F^{B\pi}=0.23$ and one can see that the overlapping region does not change significantly. We do not show the case of $F^{B\pi}=0.29$ in the figure because $B_d \to \pi^+ \pi^-$ decay would then be much larger than the experimental data. Actually, we shall see later in Table \ref{table:1} that the default value of $F^{B\pi}=0.26$ already predicts a somewhat larger value of ${\cal B}(B_d \to \pi^+ \pi^-)$ than the experimental observation. For the hard spectator parameter $X_H$, it plays an important role only in $a_2$ term of tree-dominant decays, such as $B^- \to \pi^- \pi^0$ and $B_d \to \pi^0 \pi^0$ decays. Therefore we shall not discuss further the theoretical uncertainties associated with $X_H$ in this study.

At first glance, the solution of $(\rho_A, \phi_A)=(1.6,~-45^\circ)$ seems to be in disagreement with the experimental measurements of large pure annihilation decay ${\cal B}(B_s \to \pi^+ \pi^-)$. But, a closer look at the annihilation part of the decay amplitudes in QCDF reveals that only the basic building blocks $A^i_{1,2}$ appear in $B_s \to \pi^+ \pi^-$ decay, as can be seen from Eqs. (\ref{eq:bi's},\ref{eq:amp-pipiKK}), in contrast to the case of $B_{d,s} \to \pi^\mp K^\pm$ decays which are dominated by the $A_3^f$ term.  It is a common practice, just for simplicity, to assume the annihilation parameters $\rho_A$ and $\phi_A$ to be universal for $A^i_{1,2}$ and $A_3^f$. However, as the subscript of $A_k^{i,f}$ denote different Dirac structures of effective operators, there is no a priori reason for the above assumption to be a good approximation.

Therefore it seems to be reasonable to discuss a scenario with three sets of annihilation parameters. Specifically, we introduce the parameters $(\rho_A^f, \phi_A^f)$ solely for $A_{3}^f$. As $A_3^f$ is independent of the initial state, it is justified to adopt only one set of $(\rho_A^f, \phi_A^f)$ for both $B_{u,d}$ and $B_s$ decays. In principle, these annihilation parameters may also vary mildly for different final states, due to flavor symmetry breaking effects. But we shall not discuss these effects at this stage. For $A^i_{1,2}$ terms, however, the experimental data on the branching ratios of $B_s\to \pi^+ \pi^-$ and $B_d \to K^+ K^-$ strongly prefer different annihilation parameters  between $B_s$ and $B_d$ decays, as shown in Fig. \ref{fig:pipiKK}. So we shall adopt separately $(\rho_{As}^i, \phi_{As}^i)$ for $B_s$ decays and leave $(\rho_A^i, \phi_A^i)$ just for $B_{u,d}$ decays.

\begin{figure}
\centering
\subfigure[]{
\label{fig:RhoiAs}
\includegraphics[width=0.4\textwidth]{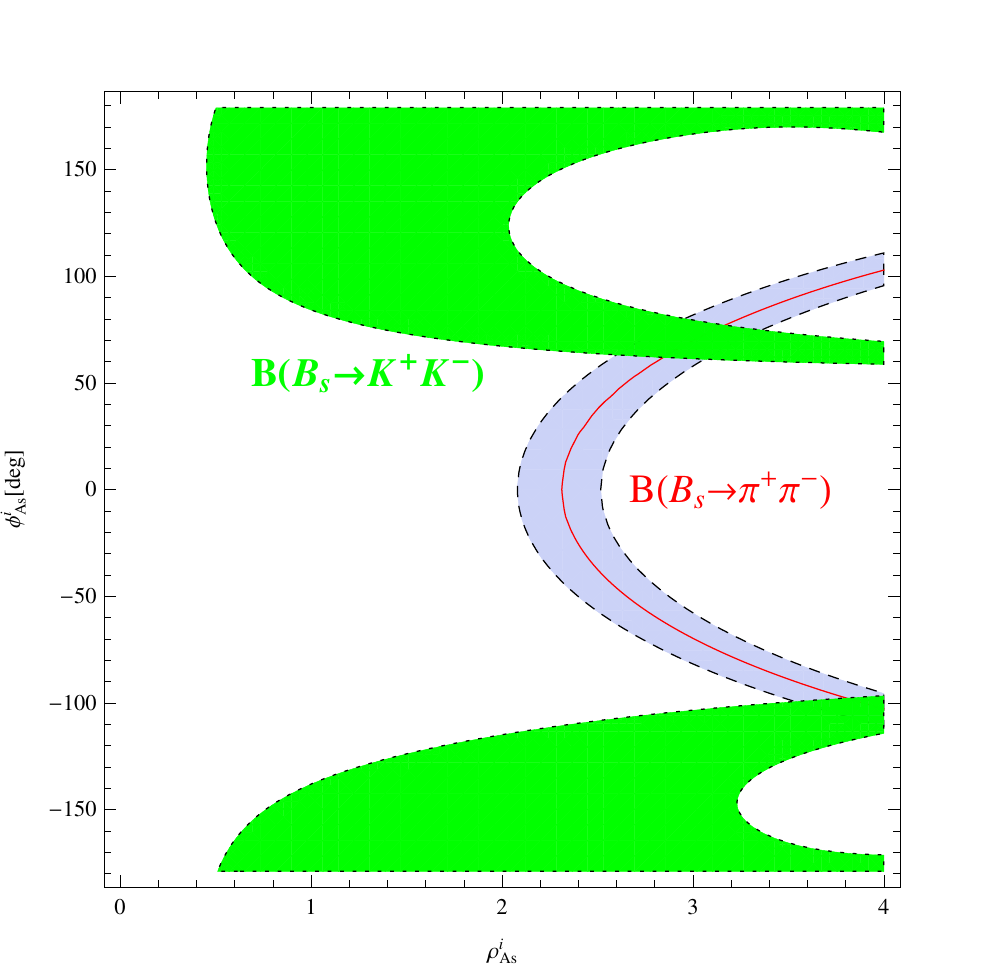}}\hspace*{1cm}
\subfigure[]{
\label{fig:RhoiA}
\includegraphics[width=0.4\textwidth]{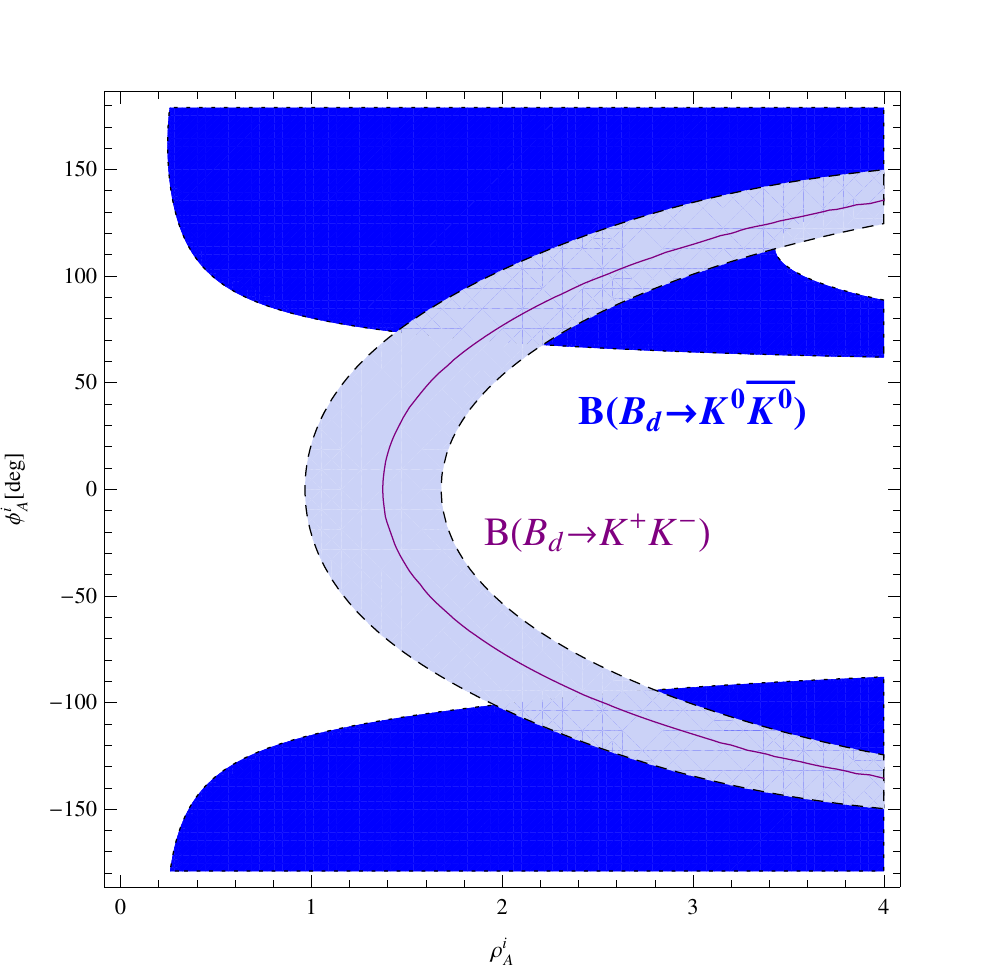}}
\caption{Left (right) figure: Contour plot of the branching ratios of $B_s \to K^+ K^-$, $\pi^+ \pi^-$ ($B_d \to K^0 \overline{K^0}$, $K^+ K^-$) decays as functions of the annihilation parameters $\rho^i_{As}$ and $\phi^i_{As}$ ($\rho^i_{A}$ and $\phi^i_{A}$). The grey (colored) regions correspond to one sigma contour with only experimental uncertainties included.}
\label{fig:Rhoi}
\end{figure}

As we have discussed earlier, $(\rho_A^f, \phi_A^f)$ can be extracted from $B_{d,s} \to \pi^\mp K^\pm$ decays to be around $(1.6,~-45^\circ)$. But to fix $(\rho_{As}^i, \phi_{As}^i)$, one has to find, besides ${\cal B}(B_s \to \pi^+ \pi^-)$, other observed $B_s$ decays which are also sensitive to the size of $A^i_{1,2}$. It turns out that currently the only candidate is $B_s \to K^+ K^-$ decay. However, its decay amplitude depends not only on $A^i_{1,2}$ but also on $A_3^f$. For convenience, we shall simply fix $\rho_A^f=1.6$, $\phi_A^f=-45^\circ$ for $A_3^f$ to estimate the parameters $\rho_{As}^i$ and $\phi_{As}^i$. The results are shown in Fig. \ref{fig:RhoiAs}. One can see that $(\rho^i_{As},~\phi^i_{As})$ are constrained to be either around $(3.0, 70^\circ)$ or around $(4.0,-100^\circ)$. However, the degeneracy can not be resolved by the direct CP asymmetry of $B_s \to K^+ K^-$, as $A_{CP}(B_s \to K^+ K^-) \simeq -10\%$ in both regions. This is because, unlike the branching ratio, $A_{CP}(B_s \to K^+ K^-)$ is numerically not sensitive to the values of $A^i_{1,2}$. Since $A^i_{1,2}$ terms play important roles only in $B_s \to \pi^+\pi^-$, $\pi^0 \pi^0$, $K^+ K^-$ and $K^0 \overline{K^0}$ decays, it is hard in practice to break the degeneracy of two solutions in $(\rho^i_{As},~\phi^i_{As})$ plane. As for $B_d$ decays, the annihilation parameters $(\rho_{A}^i, \phi_{A}^i)$ are only loosely constrained, as shown in Fig. \ref{fig:RhoiA} where $10^6{\cal B}(B_d \to K^0 \overline{K^0})=1.21 \pm 0.16$ has been used with the form factor $F^{BK}=0.30$ be adopted. Notice however that constraints of the branching ratio of $B_d \to K^0 \overline{K^0}$ in Fig. \ref{fig:RhoiA} are quite sensitive to the value of $F^{BK}$. In addition, $B_d \to K^+ K^-$ has not been observed yet. Similar to the $B_s$ case, $A^i_{1,2}$ terms are also numerically important only in $B_d \to \pi^0\pi^0$, $K^+ K^-$ and $K^0 \overline{K^0}$ decays. But hopefully more experimental data on $B_{d,s} \to \pi \rho$, $\rho \rho$ and $K^{(*)}K^{(*)}$ decays may provide a path towards better understanding on the annihilation amplitudes of charmless B decays.

With the above discussions, we may nevertheless choose a parameter scenario S1 of form factors and annihilation parameters as an illustration to show our results on full set of $\pi\pi$, $\pi K$ and $KK$ final states in Table \ref{table:1}. To be specific, the parameters in scenario S1 are chosen as follows:
\begin{align}\label{eq:S1}
&F^{B\pi}=0.26~,\hspace*{0.3cm} F^{B_s K}=0.24~,\hspace*{0.3cm}F^{BK}=0.30~,\hspace*{0.3cm}\rho_A^f=1.6~,\hspace*{0.3cm}\phi_A^f=-45^\circ~, \nonumber \\
&\rho_{As}^i=3.0~,\hspace*{0.3cm}\phi_{As}^i=70^\circ~,\hspace*{0.3cm}\rho_{Ad}^i=2.5~,\hspace*{0.3cm}\phi_{Ad}^i=100^\circ~.
\end{align}
With these parameters, we show the results on $B \to \pi\pi$, $\pi K$ and $KK$ decay modes in Table \ref{table:1}. One can see that most of the results are well consistent with the experimental measurements except $\pi \pi$ final states which are somewhat larger than the experimental data. As we have mentioned, the form factors in scenario S1 are already somewhat smaller than the estimations of QCD sum rules. But to improve our results on $B \to \pi\pi$ decays, we have tried, just for illustration, a parameter scenario S2 with even smaller form factors:
\begin{align}\label{eq:S2}
&F^{B\pi}=0.23~,\hspace*{0.3cm} F^{B_s K}=0.23~,\hspace*{0.3cm}F^{BK}=0.28~,\hspace*{0.3cm}\rho_A^f=1.7~,\hspace*{0.3cm}\phi_A^f=-40^\circ~, \nonumber \\
&\rho_{As}^i=3.4~,\hspace*{0.3cm}\phi_{As}^i=85^\circ~,\hspace*{0.3cm}\rho_{Ad}^i=2.5~,\hspace*{0.3cm}\phi_{Ad}^i=100^\circ~.
\end{align}
where the annihilation parameters have been adjusted correspondingly. In this scenario, all of the results are in good agreement with the experimental data, as shown also in Table \ref{table:1}. We do not show explicitly the result of $B_s \to \pi^0 \pi^0$ in the table as theoretically it should be exactly the same as that of $B_s \to \pi^+ \pi^-$. The color-suppressed decays $B_d \to \pi^0 \pi^0$ and $B_s \to \pi^0 K^0$ are not included in the table due to large theoretical uncertainties related to color-suppressed $a_2$ term.  For these decays, a deeper understanding is required which is however beyond the scope of this paper.
\begin{table}
\begin{tabular}{|l|c|c|l|l|c|c|l|}
\hline
~~~~~~~Mode & \multicolumn{2}{|c|}{QCDF} & Experiment & ~~~~~~~~Mode & \multicolumn{2}{|c|}{QCDF} & Experiment \\ \cline{2-3} \cline{6-7}
& S1 & S2 & & & S1 & S2 & \\
\hline
${\cal B}(B_s \to \pi^+ \pi^-)$ & $0.75$ &$0.76$ & $0.73 \pm 0.14$ &
${\cal B}(B_d \to K^+ K^-)$ & $0.12$ &$0.12$ & $0.12 \pm 0.05$ \\
\hline
${\cal B}(B_d \to \pi^- K^+)$ & $20.7$ & $20.7$ & $19.6 \pm 0.5$ &
$A_{CP}(B_d \to \pi^- K^+)$ & $-8.5$ & $-8.2$ & $-8.2 \pm 0.6$ \\
\hline
${\cal B}(B^+ \to \pi^0 K^+)$ & $13.0$ & $12.8$ & $12.9 \pm 0.5$ &
$A_{CP}(B^+ \to \pi^0 K^+)$ & $4.3$ & $3.2$ & $4.0 \pm 2.1$ \\
\hline
${\cal B}(B_d \to \pi^0 K^0)$ & $9.4$ & $9.4$ &$9.9 \pm 0.5$ &
${\cal B}(B^+ \to \pi^+ K^0)$ & $23.8$ & $23.4$ &$23.8 \pm 0.8$ \\
\hline
${\cal B}(B_s \to \pi^+ K^-)$ & $6.0$ & $5.7$ &$5.4 \pm 0.6$ &
$A_{CP}(B_s \to \pi^+ K^-)$ & $30$ & $32$ &$26 \pm 4$ \\
\hline
${\cal B}(B_s \to K^+ K^-)$ & $24.2$ & $23.8$ &$24.5 \pm 1.8$ &
$A_{CP}(B_s \to K^+ K^-)$ & $-10.2$ & $-10.3$ &  \\ \hline
${\cal B}(B_s \to K^0 \bar{K}^0)$ & $24.8$ & $24.2$ & $<66$ &
${\cal B}(B^+ \to \pi^+ \pi^0)$ & $6.5$ & $5.5$ &$5.5 \pm 0.4$ \\
\hline
${\cal B}(B_d \to \pi^+ \pi^-)$ & $6.7$ & $5.2$ &$5.1 \pm 0.2$ &
$A_{CP}(B_d \to \pi^+ \pi^-)$ & $15.9$ & $20.3$ &$29 \pm 5$ \\
\hline
${\cal B}(B^+ \to K^+ \bar{K}^0)$ & $1.5$ & $1.6$ &$1.2\pm 0.2$ &
${\cal B}(B_d \to K^0 \bar{K}^0)$ & $1.2$ & $1.3$ &$1.2 \pm 0.2$ \\
\hline
\end{tabular}
\caption{\label{table:1} CP-averaged branching ratios (in unit of $10^{-6}$) and direct CP asymmetries (in units of $10^{-2}$) of $B \to \pi\pi$, $\pi K$ and $KK$ decays in the framework of QCDF, with the input parameters of scenarios S1 and S2 given in the text.}
\end{table}

In short, we confirm in QCDF that the flavor symmetry breaking effects should be exceptionally small in $B_{d,s} \to \pi^\mp K^\pm$ decays. This is because the relevant annihilation building block $A_3^f$, as being independent of initial states, should be universal for the same final states. Therefore the measurements of these two decays could provide a robust test of the SM vs. New Physics. Nevertheless, for $B_{d,s} \to KK$ decays, the annihilation building blocks $A^i_{1,2}$ are also involved. Unfortunately, $A^i_{1,2}$ could be dependent on the initial state and the experimental data of $B_s \to \pi^+ \pi^-$ and $B_d \to K^+ K^-$ suggests that the dependence of $A^i_{1,2}$ on the initial state is likely not small. That is to say, the flavor symmetry breaking effects may be significant for the annihilation amplitudes in $B_{d,s} \to KK$ decays. As $B_{d,s} \to K^{0(*)}K^{0(*)}$ are important channels to probe the new physics effects, care must be taken to include the possibly large flavor symmetry breaking effects in future studies.

\section{Summary}
A combination of flavor symmetry and QCD factorization has suggested that $B_{d,s} \to \pi^\mp K^\pm$ and $K^{(\ast)} K^{(\ast)}$ decays could be important in testing the Standard Model and in probing new physics effects. This is partly because the flavor symmetry breaking effects between the corresponding $B_d$ and $B_s$ decays may be particularly small since the charge conjugation symmetry of the final states is respected by the final state interactions. Very recently, the first observation of direct CP violation in $B_s$ decays, $A_{CP}(B_s \to \pi^+ K^-)$, has been reported by the LHCb Collaboration which are well consistent with the flavor symmetry expectation in the Standard Model. However, the observation of pure annihilation decay $B_s \to \pi^+ \pi^-$ and the surprisingly small results on $B_d \to K^+ K^-$ appears to imply a large annihilation scenario with $\rho_A \sim 3$ in $B_s$ decays, in contrast to the case of $\rho_A \sim 1$ in $B_{u,d}$ decays in the framework of QCD factorization. This seems to indicate large flavor symmetry breaking effects between the annihilation amplitudes of $B_s$ and $B_{u,d}$ decays.  In QCD factorization, annihilation amplitudes are infrared divergent due to endpoint singularity and phenomenological parameterization has to be introduced with model dependence. For simplicity, it is a common practice, but without a priori justification, to adopt only one set of annihilation parameters for the basic building blocks $A^i_{1,2}$ and $A^f_3$ corresponding to different Dirac structures of effective four-quark operators. We notice that the annihilation amplitudes of $B_{d,s} \to \pi^\mp K^\pm$ decays are dominated by $A_3^f$ term, which is independent of the initial state except for the decay constant. Therefore the flavor symmetry breaking effects of annihilation amplitudes here must be small, in accordance with the corresponding measurements. However, only $A^i_{1,2}$ terms appear in pure annihilation decays $B_s \to \pi^+ \pi^-$ and $B_d \to K^+ K^-$. So the experimental results of pure annihilation decays strongly indicate a significant dependence of $A^i_{1,2}$ on initial and/or final states. It turns out that all of the building blocks $A^i_{1,2}$ and $A^f_3$ are involved in $B_{d,s} \to K^{(\ast)} K^{(\ast)}$ decays. Therefore when new physics effects are searched for in these decay channels, one may need to consider the potentially large flavor symmetry breaking effects in more details in the theoretical analysis.

\section*{Acknowledgement}
 KW is supported in part, by the Zhejiang University Fundamental Research Funds for the Central Universities (2011QNA3017) and the National Science Foundation of China (11245002,11275168). GZ is supported in part, by the National Science Foundation of China (11075139,11135006) and Program for New Century Excellent Talents in University (NCET-12-0480).

\end{document}